\documentclass[pre,showpacs,twocolumn,floatfix,superscriptaddress]{revtex4}
\usepackage{amssymb}

\usepackage{graphicx}
\usepackage{dcolumn}
\usepackage{bm}
\usepackage{hyperref}
\usepackage{latexsym}

\begin{document}

\title{Exact dynamics of a reaction-diffusion model with spatially
alternating rates}
\author{M.~Mobilia, B.~Schmittmann and R.~K.~P.~Zia}
\email{mmobilia, schmittm, rkpzia \, @vt.edu}
\affiliation{Center for Stochastic Processes in Science and Engineering, \\
Department of Physics, Virginia Tech, Blacksburg, VA, 24061-0435, USA}
\date{February 24, 2005}

\begin{abstract}
We present the exact solution for the full dynamics of a nonequilibrium spin
chain and its dual reaction-diffusion model, for arbitrary initial
conditions. The spin chain is driven out of equilibrium by coupling
alternating spins to two thermal baths at different temperatures. In the
reaction-diffusion model, this translates into spatially alternating rates
for particle creation and annihilation, and even negative ``temperatures''
have a perfectly natural interpretation. Observables of interest include the
magnetization, the particle density, and all correlation functions for both
models. Two generic types of time-dependence are found: if both temperatures
are positive, the magnetization, density and correlation functions decay
exponentially to their steady-state values. In contrast, if one of the
temperatures is negative, damped oscillations are observed in all
quantities. They can be traced to a subtle competition of pair creation and
annihilation on the two sublattices. We comment on the limitations of
mean-field theory and propose an experimental realization of our model in
certain conjugated polymers and linear chain compounds.
\end{abstract}

\maketitle

\section{Introduction}

Nonequilibrium many-body systems abound in the physical and life sciences
and have recently received much attention (see e.g. \cite%
{reviews,Privman,Schutzrev} and references therein). Despite these efforts,
a comprehensive theoretical framework is still lacking: As yet, there is no
equivalent of Gibbs ensemble theory for nonequilibrium systems. As a
consequence, in contrast to equilibrium statistical mechanics, macroscopic
observables cannot be computed without explicit reference to the imposed
dynamics, generally described by a master equation, and most progress in the
field is made by studying paradigmatic models \cite{Privman}. In this
context, \emph{exact} solutions of simple models are scarce, but very
precious, since they can serve as testing grounds for approximate and/or
numerical schemes and shed light on general properties of whole classes of
related models. Not surprisingly, \emph{nontrivial} solutions are almost
entirely restricted to one dimension (1D; see e.g. \cite{Privman,Schutzrev}
), and have focused on completely uniform lattices with site-independent
rates. Clearly, however, one would like to take into account more complex
situations, e.g., those with spatially varying coupling constants or rates.
Arguably, one of the simplest generalizations beyond a completely uniform
system is one with alternating rates. In the following, we consider a 1D
kinetic Ising chain (KISC), coupled to two alternating temperatures and
endowed with Glauber-like dynamics. Our analysis of this model provides a
full description of its dual counterpart, namely a reaction-diffusion system
(RDS), characterized by spatially alternating annihilation and creation
rates. Members of these two classes -- i.e., kinetic Ising and
reaction-diffusion models -- are prototypical nonequilibrium systems which
have been thoroughly studied on homogeneous lattices \cite%
{Privman,Schutzrev,FF,Glauber,Racz,Family}. Yet, they still offer surprises
and novel behaviors, when non-trivial spatial rates are investigated.

Our model was first introduced by R\`{a}cz and Zia \cite{RZ} who recognized
that (stationary) two-point correlation functions are easily found exactly,
even though spins on alternating sites are coupled to \emph{different}
temperatures. Schmittmann and Schm\"{u}ser subsequently realized that \emph{
all stationary} correlation functions are exactly calculable \cite{Beate1}.
While this information is equivalent to the full stationary solution, its
representation as $\exp (-\mathcal{H}_{\mathrm{eff}})$ is nontrivial,
involving a proliferation of longer-ranged multispin couplings \cite{Beate2}
. Finally, we recently reported the exact solution for all\emph{\ dynamic}
correlation functions, starting from a very simple initial condition, i.e.,
zero magnetization and vanishing correlations \cite{MZS}.

In this article, we complete these earlier studies by demonstrating how
competing site-dependent rates may dramatically affect the dynamics by
giving rise to an \textit{oscillatory} approach toward the nonequilibrium
steady state. We use a generating functional approach to obtain the complete
solution for all correlation functions with arbitrary initial conditions. We
focus specifically on the dynamical magnetization and the spin-spin
correlations and explore their long-time behavior. We will also consider the
dynamics of domain walls in the spin chain which can be mapped onto a
reaction-diffusion system. Interpreting our results in the language of
particle annihilation and creation, negative ``temperatures'' acquire a
natural physical meaning, leading to unexpected oscillatory dynamics. From a
more technical point of view, we are able to obtain a complete solution for
two nontrivial nonequilibrium many-body systems which provides some insight
into the solvability of two whole classes of related models.

The mapping to a reaction-diffusion system is of interest for two reasons.
On the theoretical side, the equations for densities and correlation
functions in the RDS form an \emph{infinite hierarchy} whose solution is not
at all apparent until one recognizes the equivalent spin chain model. Also,
from an experimental perspective, it is well known that diffusion-limited
reactions with annihilation and creation of pairs of particles are good
models for the photogrowth properties of excited states
(solitons/antisoliton pairs) in certain conjugated polymers and linear chain
compounds \cite{exp1,exp2,exp3}. We propose that spatially alternating
creation/annihilation rates in these systems -- especially in MX chain
compounds -- can be generated with the help of a laser with spatially
modulated power output.

This article is organized as follows: In the next section we introduce the
kinetic spin chain and its dual reaction-diffusion system. Section III
presents the complete solution of the spin chain. Some technical details are
relegated to two Appendices. In Section IV, we map the two-temperature spin
chain onto a reaction-diffusion system with alternating rates, whose density
and correlation functions are computed. We analyze the conditions under
which damped oscillations characterize the approach to the steady state, and
we compare our exact results to a simple mean-field description. Section V
is devoted to a brief discussion of the solvability of related models, with
Section VI reserved for our conclusions.

\section{The models}

We consider two closely related nonequilibrium many-particle systems on a
one-dimensional lattice: (\emph{i}) a kinetic Ising spin chain (KISC)
endowed with a generalized Glauber-like dynamics; and (\emph{ii}) a
reaction-diffusion system (RDS), with spatially periodic pair annihilation
and creation rates. For convenience, we restrict ourselves to a periodic
lattice (a ring) with an even number of sites and study the thermodynamic
limit. We expect our exact results to be valid for the general cases of odd
number of sites and/or arbitrary boundary conditions, apart from the usual
caveats.

Since the RDS follows from the spin chain via a duality relationship, we
focus mainly on the detailed description of model (\emph{i}). A spin
variable, $\sigma _{j}=\pm 1$, denotes the value of the spin at site $j$,
with $j=1,2,...L$, and $L$ an even integer. Nearest-neighbor spins interact
according to the usual Ising Hamiltonian: $\mathcal{H}=-J\sum_{j}\sigma
_{j}\sigma _{j+1}$, where $J>0$ ($J<0$) is the (anti-) ferromagnetic
exchange coupling. Our model is endowed with a \emph{nonequilibrium}
generalization of the usual Glauber \cite{Glauber} dynamics: spins on even
and odd sites experience different temperatures, $T_{e}$ and $T_{o}$, which
implies the violation of detailed balance \cite{RZ,Beate1,Beate2}. To be
specific, a configuration $\{\sigma _{1},\sigma _{2},\dots ,\sigma _{L}\}$
evolves into a new one by random sequential spin flips: A spin $\sigma _{j}$
flips to $-\sigma _{j}$ with rate
\begin{eqnarray}
w_{j}(\{\sigma \}) &\equiv &w_{j}(\sigma _{j}\rightarrow -\sigma _{j})
\nonumber \\
&=&\frac{1}{2}-\frac{\gamma _{j}}{4}\sigma _{j}\left( \sigma _{j-1}+\sigma
_{j+1}\right)   \label{rates}
\end{eqnarray}
where $\gamma _{2i}=\gamma _{e}=\mathrm{tanh}\left( 2J/k_{b}T_{e}\right) $
and $\gamma _{2i+1}=\gamma _{o}=\mathrm{tanh}\left( 2J/k_{b}T_{o}\right) $,
on even ($j=2i$) and odd ($j=2i+1$) sites. The time-dependent probability
distribution $P(\{\sigma \},t)$ obeys the master equation:
\begin{eqnarray}
&&\partial _{t}P(\{\sigma \},t)=  \nonumber \\
&=&\sum_{j}\left[ w_{j}(\{\sigma \}^{j})P(\{\sigma \}^{j},t)-w_{j}(\{\sigma
\})P(\{\sigma \},t)\right]   \label{meq}
\end{eqnarray}
where the state $\{\sigma \}^{j}$ differs from $\{\sigma \}$ only by the
spin flip of $\sigma _{j}$. Our main goal in this work is to compute the
time-dependent distribution $P(\{\sigma \},t)$. To do so, we compute \textit{%
all} correlation functions $\langle \sigma _{j_{1}}\dots \sigma
_{j_{n}}\rangle _{t}\equiv \sum_{\{\sigma \}}\sigma _{j_{1}}\dots \sigma
_{j_{n}}P(\{\sigma \},t)$ and invoke the following relationship \cite%
{Glauber}:
\begin{eqnarray}
2^{L}\,P(\{\sigma \},t)=1 &+&\sum_{i}\sigma _{i}\langle \sigma _{i}\rangle
_{t}+\sum_{j>k}\sigma _{j}\sigma _{k}\langle \sigma _{j}\sigma _{k}\rangle
_{t}+  \nonumber  \label{rel} \\
&+&\sum_{j>k>l}\sigma _{j}\sigma _{k}\sigma _{l}\langle \sigma _{j}\sigma
_{k}\sigma _{l}\rangle _{t}+...
\end{eqnarray}%
This expression illustrates that the knowledge of \emph{all} equal-time
correlation functions is equivalent to the complete knowledge of the
distribution function $P(\{\sigma \},t)$. Recently, this implication was
exploited for the steady state \cite{Beate1}, and for the time-dependent
situation yet restricted to a particularly simple initial condition \cite%
{MZS}.

The spin-flip dynamics of this Ising chain can be expressed in terms of the
creation, annihilation and diffusion of \emph{domain walls}, i.e., pairs of
spins with opposite sign. For example, flipping $\sigma _{j}$ in the local
configuration $\sigma _{j-1}=\sigma _{j}=\sigma _{j+1}=+1$ creates two
domain walls: $\sigma _{j-1}=-\sigma _{j}$ and $\sigma _{j}=-\sigma _{j+1}$,
located on the \emph{bonds} $(j-1,j)$ and $(j,j+1)$. Similarly, flipping $
\sigma _{j}$ in the local configuration $\sigma _{j-1}=\sigma _{j}=-\sigma
_{j+1}=+1$ has the effect of moving the domain wall on bond $(j,j+1)$ by one
lattice constant to the left, corresponding to domain wall diffusion. By
identifying a domain wall with a ``particle'', $A$, our spin-flip dynamics
can be recast as a reaction-diffusion model, and the two examples translate
into $\emptyset \emptyset \rightarrow AA$ and $\emptyset A\rightarrow
A\emptyset $, respectively. The mapping from the KISC into its dual RDS is
described in detail in Table 1.
\begin{table*}[!t]
\caption{Basic processes underlying the KISC (left) and RDS (middle)
dynamics }%
\begin{tabular}{|c|c|c|}
\hline
\textit{Spin flip of site $j$} & \textit{Reactions at bonds next to site $j$}
& \textit{Rates} \\ \hline
$+ - - \longrightarrow + + -$ and $- - + \longrightarrow - + +$ & $%
A\emptyset \longrightarrow \emptyset A$ and $\emptyset A \longrightarrow A
\emptyset $ & $1/2 $ \\ \hline
$+ - + \longrightarrow + + +\;$ ($j$ even) & $A A \longrightarrow \emptyset
\emptyset\;$ ($j$ even) & $(1 +\gamma_e)/2\;$ \\ \hline
$+ - + \longrightarrow + + +$ ($j$ odd) & $A A \longrightarrow \emptyset
\emptyset \;$ ($j$ odd) & $(1 +\gamma_o)/2 $ \\ \hline
$+ + + \longrightarrow + - +$ ($j$ even) & $\emptyset \emptyset
\longrightarrow A A \;$ ($j$ even) & $(1-\gamma_{e})/2$ \\ \hline
$+ + + \longrightarrow + - +$ ($j$ odd) & $\emptyset \emptyset
\longrightarrow A A \;$ ($j$ odd) & $(1 -\gamma_{o})/2$ \\ \hline
\end{tabular}
\end{table*}

Clearly, the presence of alternating temperatures $T_{e}$, $T_{o}$ in spin
language translates into alternating pair annihilation and creation rates $%
(1\pm \gamma _{e,o})/2$ in the RDS. We can see easily that letting $T_{e}$
or $T_{o}$ vanish simply prohibits pair creation entirely at even or odd
sites. Remarkably, we can derive an additional, and possibly rather
unexpected, benefit from this mapping: Assigning \emph{negative} values for
the temperatures $T_{e}$ and/or $T_{o}$ may appear artificial in the KISC,
but is \emph{perfectly natural} in the RDS: For example, $T_{e}<0$ simply
corresponds to a creation rate $(1-\gamma _{e})/2>1/2$ which is easily
implemented in a simulation. In other words, the RDS version is physically
meaningful, and readily accessible, on a much wider parameter space.

\section{Complete solution of the kinetic spin chain}

In this section, we completely solve the dynamics of the KISC. It was shown
previously \cite{Aliev} that the generating function, and hence the full
distribution $P(\{\sigma \},t)$, of a broad class of Ising models can be
computed from two very basic observables, namely:\ (\emph{i})\emph{\ }the
magnetization, $m_{j}(t)=\langle \sigma _{j}\rangle _{t}$ for \emph{arbitrary%
} initial condition, and (\emph{ii})\emph{\ }a\emph{\ particular }two-point
equal-time correlation function, $c_{j,k}(t)=\langle \sigma _{j}\sigma
_{k}\rangle _{t}$, the resultant from the special initial conditions: $%
m_{j}(0)=c_{j,k}(0)=0$ $\ $(see Appendix A for a more detailed discussion of
this statement). Here, $\langle \cdot \rangle _{t}\equiv $ $\sum_{\{\sigma
\}}\cdot \,P(\{\sigma \},t)$ denotes the usual configurational average. In
the following, we assemble the necessary information about these two
observables.

\subsection{The general $t$-dependent magnetization.}

From our earlier work \cite{MZS}, we recall that the magnetization $%
m_{j}(t)=\langle \sigma _{j}\rangle _{t}$ of the KISC obeys the equation of
motion,
$
\frac{d}{dt}m_{j}(t)=\frac{\gamma _{j}}{2}\left[ m_{j-1}(t)+m_{j+1}(t)\right]
-m_{j}(t)
$
which is easily derived from the master equation, Eqn. (\ref{meq}). As shown in 
\cite{MZS}, the general
solution of this linear equation takes the form
$
m_{j}(t)=\sum_{k}M_{j,k}\left( t\right) m_{k}(0), $
where the ``propagator''$M_{j,k}\left( t\right) $ can be written in term of
modified Bessel functions of first kind $I_{n}(t)$ \cite{Abramowitz}:
\begin{equation}
M_{j,k}\left( t\right) =e^{-t}\sqrt{\frac{\gamma _{j}}{\gamma _{k}}}
I_{k-j}(\alpha t)  \label{M(t)}, \; \text{with} \;\alpha \equiv 
(\gamma _{e}\gamma _{o})^{1/2}\;
\end{equation}

If $\gamma _{e}\gamma _{o}<0$, the propagator becomes $M_{j,k}\left(
t\right) =i(-1)^{(k-j)/2}\,|\gamma _{j}/\gamma
_{k}|^{1/2}e^{-t}J_{k-j}(|\alpha |t)$ \cite{MZS}, where $J_{n}(t)$ is a
Bessel function of the first kind, with damped oscillatory asymptotic
behavior \cite{Abramowitz}. This translates into an oscillatory decay of the
magnetization \cite{MZS}.

\subsection{A special two-point equal-time correlation function.}

The second fundamental quantity, i.e., the equal-time spin-spin correlation
function $c_{k,j}(t)$, with $k>j$, is already known from \cite{MZS}. For our
purposes, it suffices to consider an initial condition with zero
magnetization and zero initial correlations. With the boundary condition $%
\langle \sigma _{j}\sigma _{k}\rangle _{t}=1$ for $j=k$, this basic
correlation depends only on the distance between the two sites and their
parity, $\mu (k),\mu (j)\in \{e,o\}$ \cite{MZS}:
\begin{eqnarray}
c_{k,j}(t) &\equiv &c_{k-j\,}^{\mu (k),\mu (j)}(t)  \nonumber \\
&=&\frac{\bar{\gamma}}{\alpha ^{2}}\sqrt{\gamma _{j}\gamma _{k}}%
\,\,(k-j)\,\int_{0}^{2t}\frac{d\tau }{\tau }\,e^{-\tau }\,I_{k-j}(\alpha
\tau )  \label{two-spin}
\end{eqnarray}
where
\begin{equation}
\bar{\gamma}\equiv (\gamma _{e}+\gamma _{o})/2.  \label{gammbar}
\end{equation}
For long times, these settle into their stationary values \cite{RZ,Beate1},
independent of initial conditions:
\begin{equation}
\langle \sigma _{j}\sigma _{k}\rangle _{\infty }\equiv c_{k,j}(\infty )=\frac{\bar{\gamma}}{\sqrt{\gamma _{j-1}\gamma _{k-1}}}\,\omega ^{k-j},
\label{cinf}
\end{equation}
where
\begin{equation}
\omega \equiv \frac{\alpha }{1+\sqrt{1-\alpha ^{2}}},  \label{omega0}
\end{equation}
a quantity that reduces to the familiar $\tanh \left( J/k_{b}T\right) $ in the
equilibrium Ising chain. The approach to these values is exponential and
monotonic, as $e^{-2(1-\alpha )t}t^{-3/2}$, provided $\gamma _{e}\gamma
_{o}>0$. However, for $\gamma _{e}\gamma _{o}<0$, the approach is
oscillatory and damped by $e^{-2t}t^{-3/2}$ \cite{MZS}. For later reference,
it is convenient to display the parity dependence explicitly. Since
translation invariance ensures $c_{k-j\,}^{oe}(t)=c_{k-j\,}^{eo}(t)$, we
need to distinguish three types of correlations. The simplest display, which
manifestly shows the underlying symmetries, is
\begin{equation}
\left(
\begin{array}{c}
c_{k-j\,}^{ee}(t) \\
c_{k-j\,}^{eo}(t) \\
c_{k-j\,}^{oo}(t)
\end{array}
\right) =\left(
\begin{array}{c}
{\bar{\gamma}/}\gamma _{o} \\
{\bar{\gamma}/}\alpha  \\
{\bar{\gamma}/}\gamma _{e}
\end{array}
\right) (k-j)\,\int_{0}^{2t}\frac{d\tau }{\tau }e^{-\tau }I_{k-j\,}(\alpha
\tau ).  \label{FINAL}
\end{equation}
Note that the last factor is of exactly the same form as in the ordinary
Ising chain coupled to a single thermal bath, the only difference being the
geometric mean of the two $\gamma $'s here plays the role of $\gamma =\tanh
\left( 2J/k_{b}T\right) $. Before turning to the general case, let us remind
the reader that Eqns. (\ref{two-spin}) and (\ref{FINAL}) give the
time-dependent correlations only for a system with no initial magnetization
and two-spin correlations (e.g., a random distribution). In particular,
these forms, also used in the next sections, should not be confused with the more general cases considered in Appendix B.

\subsection{Generating function and general multi-spin correlations.}

In this section, starting from our knowledge of $m_{j}(t)$ and $c_{k,j}(t)$,
we compute the generating function of the KISC, following \cite{Aliev}. By
construction, this generating function allows us to find\emph{\ all}
correlation functions, subject to \emph{arbitrary }initial conditions. A few
additional technical details are provided in Appendix A.

The generating function is defined via $\Psi (\{\eta \},t)\equiv \langle
\prod_{j}\left( 1+\eta _{j}\sigma _{j}\right) \rangle _{t}$, where the $
\left\{ \eta _{j}\right\} $ are standard Grassmann variables \cite{Aliev,Itz}%
. In the thermodynamic limit, $L\rightarrow \infty $, it simplifies to
\begin{eqnarray}  \label{gen}
\Psi(\{\eta\},t) &=& \left\langle \prod_{j}\left(1+\sigma_j \sum_{k}\eta_k
M_{k,j}(t)\right)\right\rangle_0  \nonumber \\
&\times& \mathrm{exp}\left(\sum_{j_2>j_1} \eta_{j_1} \eta_{j_2} \,
c_{j_2,j_1}(t)\right),
\end{eqnarray}
If the initial magnetization and all initial correlations vanish, the
average $\langle ...\rangle _{0}$ on the right hand side of Eqn. (\ref{gen})
reduces to unity, and one recovers the bilinear form for $\Psi (\{\eta \},t)$
which we already reported in \cite{MZS}. Eqn. (\ref{gen}) is one of the key
results of this paper.

Given the generating function, all correlation functions can be obtained by
simple differentiation \cite{MZS,Aliev}: 
$\langle \sigma _{j_{1}}\dots \sigma _{j_{n}}\rangle _{t}= \left. \frac{\partial
^{n}\Psi (\{\eta \},t)}{\partial \eta _{j_{n}}\dots \partial \eta _{j_{1}}}
\right|_{\{\eta \}=0}$. As an illustration, we compute the equal-time spin-spin correlation
functions, for $k>j$: 
\begin{eqnarray}
&&\langle \sigma _{j}\sigma _{k}\rangle _{t}=\left. \frac{\partial ^{2}\Psi
(\{\eta \},t)}{\partial \eta _{k}\partial \eta _{j}} \right| _{\{\eta
\}=0}=c_{k,j}(t)+  \label{2P} \\
&+&\sum_{\ell <m}\langle \sigma _{\ell }\sigma _{m}\rangle _{0}\left[
M_{\ell ,j}(t)M_{m,k}(t)-M_{\ell ,k}(t)M_{m,j}(t)\right]   \nonumber
\end{eqnarray}%
We emphasize that this is a completely \emph{general result}, valid for 
\emph{any} initial conditions, whether homogeneous or inhomogeneous,
translationally invariant or not. The two terms in (\ref{2P}) have simple
interpretations. While the second term reflects the decay of the \emph{%
initial }correlations, the first provides the buildup to the final
stationary values given above (\ref{cinf}). Thus, we see explicitly how the
stationary spin-spin correlation function becomes independent of the initial
values. 

Higher order correlations are can also be evaluated but are rather complex
for general initial conditions. For uncorrelated, non-magnetized initial
conditions, however, they simplify significantly \cite{MZS}. For example,
the $4$-point function $\langle \sigma _{j_{1}}\sigma _{j_{2}}\sigma
_{j_{3}}\sigma _{j_{4}}\rangle _{t}$ factorizes into two-point functions,
according to 
$\langle \sigma _{j_{1}}\sigma _{j_{2}}\sigma _{j_{3}}\sigma _{j_{4}}\rangle
_{t}
=c_{j_{2},j_{1}}(t)c_{j_{4},j_{3}}(t)-c_{j_{3},j_{1}}(t)c_{j_{4},j_{2}}(t)
+c_{j_{4},j_{1}}(t)c_{j_{3},j_{2}}(t)$
for $j_{4}\geq j_{3}\geq j_{2}\geq j_{1}$ \cite{MZS}. Similar factorizations hold for
all correlations. Their steady-state behavior can be computed directly from
the master equation \cite{Beate1} or from the stationary limit of the
generating function, $\Psi (\{\eta \},\infty )=\mathrm{exp}\left(
\sum_{k>j}\eta _{j}\eta _{k}c_{k,j}(\infty )\right) $. Thanks to this simple
form, the $2n$-point correlations factorize into a product of 2-point
correlations:\ $\langle \sigma _{j_{1}}\sigma _{j_{2}}\dots \sigma
_{j_{2n-1}}\sigma _{j_{2n}}\rangle _{\infty }=\langle \sigma _{j_{1}}\sigma
_{j_{2}}\rangle _{\infty }\dots \langle \sigma _{j_{2n-1}}\sigma
_{j_{2n}}\rangle _{\infty }$, where $j_{2n}>j_{2n-1}>\dots >j_{2}>j_{1}$.

Finally, following Refs \cite{Glauber,MZS}, we can also derive the \emph{\
unequal}-time spin-spin correlation functions $c_{k,j}(t^{\prime };t)$
describing how a spin on site $k$ at time $t$ is correlated with the spin on
site $j$ at a later time $t+t^{\prime }$: 
\begin{widetext}
\begin{eqnarray}
c_{k,j}(t^{\prime };t) &=& \sum_{\ell }M_{j\ell }\left( t^{\prime }\right)
\langle \sigma _{k}\sigma _{\ell}\rangle _{t} \nonumber \\
&=&\sum_{\ell}M_{j,\ell}(t') c_{k,\ell}(t)  
+\sum_{\ell}\sum_{k_1<\ell_1}\langle \sigma_{k_1}\sigma_{\ell_1}\rangle_{0} M_{j,\ell}(t')
\left[
M_{k_1,k}(t)M_{\ell_1,\ell}(t) - M_{k_1,\ell}(t)M_{\ell_1,k}(t)
\right]
\end{eqnarray}
\end{widetext} 
As an illustration of these general results, in Appendix B we specifically
compute the spin-spin correlation functions for general translationally
invariant initial conditions.

\section{Consequences for a reaction-diffusion model with alternating rates}

In this section, our exact results will be translated into the language of
the corresponding reaction-diffusion model. We first associate a site $\hat{
\jmath}$ on the dual lattice with every bond ($j-1,j$) of the original
chain. Since the particles of the RDS are identified with domain walls in
the spin chain, they obviously reside on the dual lattice. Each site $\hat{%
\jmath}$ can be occupied by at most one particle, described by an occupation
variable $n_{\hat{\jmath}}$ which takes the value $0$ ($1$) if the site is
empty (occupied). Since a domain wall involves two neighboring spins, the
mapping from spin to particle language is nonlinear, namely, $n_{\hat{\jmath}%
}=\frac{ 1}{2}\left[ 1-\sigma _{j-1}\sigma _{j}\right] $. As before, we seek
the probability, $\hat{P}(\{n\},t)$, to find configuration $\{n\}$ at time $t
$, and its averages:\ the local particle density $\rho _{\hat{\jmath}%
}(t)\equiv \left\langle n_{\hat{\jmath}}\right\rangle _{t}\equiv $ $%
\sum_{\{n\}}n_{\hat{ \jmath}}\hat{P}(\{n\},t)$ and the $m$-point correlation
functions, $\langle n_{\hat{\jmath}_{1}}\dots n_{\hat{\jmath}_{m}}\rangle
_{t}\equiv \sum_{\{n\}}n_{\hat{\jmath}_{1}}\dots n_{\hat{\jmath}_{m}}\hat{P}%
(\{n\},t)$. To simplify notation, we continue to denote averages by $%
\left\langle \cdot \right\rangle _{t}$ for both spins and occupation
variables, even though they are controlled by different statistical weights, 
$P(\{\sigma \},t)$ and $\hat{P}(\{n\},t)$, respectively. In each case, it
should be perfectly clear from the context which distribution is relevant.
The dynamics of our model is characterized by symmetric diffusion of
particles (with rate $1/2$) and pair annihilation/creation of particles with
spatially alternating rates $ (1\pm \gamma _{j})/2$. In this case, the two
particles are created on the (dual lattice)\ sites $\hat{\jmath}$ and $\hat{%
\jmath}+1$, by flipping a spin on the (original lattice)\ site $j$. Since $%
\gamma _{j}$ can be positive or negative, subject only to $-1\leq \gamma
_{j}\leq 1$ for all $j$ , two very distinct behaviors emerge: (\emph{i})
when both $\gamma _{e}$ and $\gamma _{o}$ are positive (corresponding to
positive ``temperatures''$\;$in the spin model), the annihilation process
always occurs with a \emph{larger} rate than the creation process,
irrespective of whether $j$ is even or odd; (\emph{ii}) when, e.g., $\gamma
_{o}$ is negative and $\gamma _{e}$ positive, the system displays a \textit{%
mild site-dependent frustration}:\ at even sites $j$ (i.e., $\hat{\jmath}$
even and $\hat{\jmath}+1$ odd) annihilation is more likely than creation,
whereas the situation is reversed on odd sites (where $\hat{\jmath}$ odd and 
$\hat{\jmath}+1$ even). As we will see shortly, this gives rise to \textit{%
oscillatory} dynamics.

Before diving into the details, some further remarks on physical
realizations of this model are in order. When the rates are uniform ($\gamma
_{e}=\gamma _{o}$), it is well known that such an RDS describes the dynamics
of photo-excited solitons in conjugated polymers or linear chain compounds.
MX chain compounds, [Pt($en$)$_{2}$][Pt($en$)Cl$_{2}$]$Y_{4}$, where $Y$
stands for ClO$_{4}$ or BF$_{4}$ and $(en)$ for enthylenediamine, are of
particular experimental interest \cite{exp1,exp2}. In these compounds,
photogenerated solitons are so long-lived that they can be experimentally
studied. Irradiation with continuous wave (non-pulsed) blue light generates
soliton-antisoliton pairs which can diffuse apart or annihilate. Their
static and dynamic properties are in quantitative agreement with theoretical
models \cite{FF,MM}. Since creation, annihilation, and hopping rates can be
controlled by tuning the laser power, we believe that spatially alternating
rates such as ours will be generated if an MX chain compound is exposed to a
spatially modulated light intensity.

Returning to our model, our goal in this section is first, to derive all
correlation functions from our exact solution of the KISC. We will also
comment on the validity of a simple mean-field theory which is widely used
for the homogeneous ($\gamma _{e}=\gamma _{o}$)\ case \cite{Mendoca,MM}.
Further, we show that particle hops in the RDS develop a peculiar
directional preference in the steady state, even though there is no explicit
bias in the rates, boundary or initial conditions. Finally, we illustrate
how oscillatory behaviors may result from a competition of the underlying
processes.

\subsection{Density of particles in the RDS}

The observable of most immediate interest is the average density of
particles, $\rho _{\hat{\jmath}}(t)$, in the RDS. Its equation of motion can
be derived easily from the associated master equation, resulting in: 
\begin{eqnarray}
2\frac{d}{dt}\rho _{\hat{\jmath}}(t) &=&(2-\gamma _{j}-\gamma
_{j-1})+(\gamma _{j-1}\rho _{\hat{\jmath}-1}(t)  \nonumber  \label{EOMconc}
\\
&+&\gamma _{j}\rho _{\hat{\jmath}+1}(t))-(4-\gamma _{j}-\gamma _{j-1})\rho _{%
\hat{\jmath}}(t)  \nonumber \\
&-&2\left[ \gamma _{j}\langle n_{\hat{\jmath}}n_{\hat{\jmath}-1}\rangle
_{t}+\gamma _{j+1}\langle n_{\hat{\jmath}}n_{\hat{\jmath}+1}\rangle _{t}
\right] 
\end{eqnarray}
It is worthwhile noting that this equation is the first member of an
infinite hierarchy, connecting lower-order correlations to higher-order
ones. In general, such hierarchies cannot be solved directly, without
recourse to crude approximations. Here, the mapping to the spin chain
develops its full power, allowing us to compute all correlation functions
for the RDS.

The mapping from spins to particles implies that $\rho _{\hat{\jmath}
}(t)\equiv \left\langle n_{\hat{\jmath}}\right\rangle =\frac{1}{2}
[1-\left\langle \sigma _{j-1}\sigma _{j}\right\rangle _{t}]$, so that we can
just turn to Eqn. (\ref{2P}) to read off the answer. To express it fully in
RDS language, we also need to translate the initial correlations, $\langle
\sigma _{k}\sigma _{\ell }\rangle _{0}$. For $k<\ell $ and any $t$
(including $t=0$), we may write $\langle \sigma _{k}\sigma _{\ell }\rangle
_{t}=\langle \sigma _{k}\sigma _{k+1}\sigma _{k+1}\sigma _{k+2}...\sigma
_{\ell -1}\sigma _{\ell }\rangle _{t}=\langle (1-2n_{\hat{k}+1})(1-2n_{\hat{k
}+2})\dots (1-2n_{\hat{\ell}})\rangle _{t}$ \cite{Santos,MM} whence we
obtain, for arbitrary initial condition: 
\begin{eqnarray}
\rho _{\hat{\jmath}}(t) &=&\frac{1}{2}\left\{ 1-c_{j,j-1}(t)\right\}  
\nonumber  \label{dens} \\
&-&\frac{1}{2}\sum_{\hat{k}<{\hat{\ell}}}\langle (1-2n_{\hat{k}+1})(1-2n_{
\hat{k}+2})\dots (1-2n_{{\hat{\ell}}})\rangle _{0}  \nonumber \\
&\times &\left[ M_{k,j-1}(t)M_{\ell ,j}(t)-M_{k,j}(t)M_{\ell ,j-1}(t)\right] 
\end{eqnarray}
Since the ``propagators''\ $M_{i,j}(t)$ decay exponentially as $t\rightarrow
\infty $, the steady-state density is independent of initial conditions and
spatially uniform:\ \ 
\begin{equation}
\rho (\infty )\equiv \rho _{j}(\infty )=\frac{1}{2}\left(1-\frac{{\bar{
\gamma}}}{\sqrt{\gamma _{e}\gamma _{o}}}\;\omega \right) .
\label{densstat}
\end{equation}

In Appendix B, we explicitly evaluate Eqn. (\ref{dens}) for a generic but
simple initial condition, characterized by a uniform, uncorrelated initial
distribution of particles, with density $\rho (0)$. For simplicity, we
discuss only its long-time limit here, for $\rho (0)=1/2$. We observe two
distinct kinds of behaviors:

\noindent (\emph{i}) When $\gamma _{e}\gamma _{o}>0$, the stationary density
of particles is approached exponentially fast [except when $\gamma_e=\gamma_o=\pm 1$, see
 (\ref{rhocrit})], with inverse relaxation-time $
2(1-\alpha )$ , and a subdominant power-law prefactor $t^{-3/2}$: 
\begin{eqnarray}
\rho (t) &=&\frac{1}{2}\left( 1-\frac{{\bar{\gamma}}}{\alpha }
\,\int_{0}^{2t} \frac{d\tau }{\tau }e^{-\tau }I_{1}(\alpha \tau )\right) 
\nonumber \\
&\simeq &\rho (\infty )+\frac{t^{-3/2}e^{-2(1-\alpha )t}}{2\sqrt{2\pi \alpha 
}(1-\alpha )}.  \label{dens_ex1}
\end{eqnarray}
This long-time behavior is very similar to that found in the usual ($\gamma
_{e}=\gamma _{o}\neq \pm 1$) pair diffusion, annihilation, and creation process $
AA\rightleftarrows \emptyset \emptyset $ \cite{FF,MM}.

\noindent (\emph{ii}) For $\gamma _{o}\gamma _{e}<0$, we observe a
competition between the different processes. For example, when $-1\leq
\gamma _{o}<0$ and $0<\gamma _{e}\leq 1$, the annihilation (creation)
reaction dominates on even (odd) sites. As a result, the stationary density
is reached exponentially fast with \emph{damped oscillations}: 
\begin{eqnarray}
&& \rho (t)=\frac{1}{2}\left( 1-\frac{{\bar{\gamma}}}{|\alpha |}
\,\int_{0}^{2t}\frac{d\tau }{\tau }e^{-\tau }J_{1}(\alpha \tau )\right)
\label{dens_ex2} \\
&& \simeq \rho (\infty ) -e^{-2t}\,\left[ \frac{\sin {(2|\alpha |t-\frac{\pi 
}{4})}+|\alpha |\cos {(2|\alpha |t-\frac{\pi }{4})}}{4(1+|\alpha |^{2})t 
\sqrt{\pi |\alpha |t}}\right]  \nonumber
\end{eqnarray}

For initial densities other than $1/2$, as shown in Appendix B, only the
amplitude, or the subdominant power-law prefactor, of the expressions (\ref%
{dens_ex1}, \ref{dens_ex2}) change. Since they depend on all parameters of
the model, including the initial density, the dynamics is manifestly \textit{%
\ nonuniversal}.

\subsection{Two-point correlation functions of the RDS}

A deeper understanding of the time-dependent spatial structures of our RDS
is provided by the $m$-point correlation functions, $\langle n_{\hat{\jmath}
_{1}}\dots n_{\hat{\jmath}_{m}}\rangle _{t}$ of such a model. These are
related to the correlation functions of the dual spin chain, via $\langle n_{%
\hat{\jmath}_{1}}\dots n_{\hat{\jmath}_{m}}\rangle _{t}$ $=2^{-m}\langle
(1-\sigma _{j_{1}-1}\sigma _{j_{1}})\dots (1-\sigma _{j_{m}-1}\sigma
_{j_{m}})\rangle _{t}$, and are therefore exactly known. It is interesting
to note that the $m$-point correlation function for the RDS is a
superposition of all $2n$-point correlation functions for the spin chain,
with $n=1,2,...,m$. In the following, we focus on the most directly
observable correlation, namely, the two-point function. To avoid unnecessary
technical complications which add little insight, we specifically consider a
system that is initially homogeneously half-filled with $A$ particles,
without any initial correlations:\ $\rho _{\hat{\jmath}}(0)=1/2$ and $
\langle n_{\hat{\jmath}}(0)n_{\hat{k}}(0)\rangle =1/4$, for $\hat{\jmath}
\neq \hat{k}$. Such an initial configuration corresponds, in the KISC
picture, to a system with initially neither magnetization nor correlations.
In this case, as we showed in \cite{MZS}, the generating function takes a
rather simple \textit{bilinear} form which simplifies the spin-spin
correlations.

With this initial condition, both the spin chain and the RDS are
translationally invariant, modulo period 2. As a result, the two-point
correlations $\mathcal{C}_{\hat{k}-\hat{\jmath}}^{\mu (\hat{k}),\mu (\hat{
\jmath})}(t)\equiv \langle n_{\hat{\jmath}}n_{\hat{k}}\rangle _{t}$ between
two sites $\hat{\jmath}$ and $\hat{k}$ (with $\hat{k}>\hat{\jmath}$) depend
only on the distance $\hat{k}-\hat{\jmath}$ and the parity $\mu (\hat{k}
),\mu (\hat{\jmath})\in \{e,o\}$ of the two sites. We therefore need to
distinguish four distinct correlation functions: $\mathcal{C}_{\hat{k}-\hat{
\jmath}}^{ee}(t)$, $\mathcal{C}_{\hat{k}-\hat{\jmath}}^{eo}(t)$, $\mathcal{C}
_{\hat{k}-\hat{\jmath}}^{oe}(t)$, and $\mathcal{C}_{\hat{k}-\hat{\jmath}
}^{oo}(t)$. By virtue of our mapping to the KISC, these are determined by
the $2$- and $4$-point \emph{spin} correlations as explained in Section III.C
[from Eqns. (\ref
{FINAL}) and (\ref{gen})]. Exploiting
translational invariance, the two-point correlations for the RDS, for $\hat{
k
}>\hat{\jmath}$, then follow as: 
\begin{eqnarray}
\langle n_{\hat{\jmath}}n_{\hat{k}}\rangle _{t}&=&\frac{1}{4}\left[ \left(
1-c_{1}^{eo}(t)\right) ^{2}-\langle \sigma _{j}\sigma _{k}\rangle _{t}^{2} 
\right]  \nonumber \\
&+&\frac{1}{4}\langle \sigma _{j-1}\sigma _{k}\rangle _{t}\langle \sigma
_{j}\sigma _{k-1}\rangle _{t},  \label{corrRD2P}
\end{eqnarray}

Now we are ready to discuss our results. First of all, we consider a special
case, namely, nearest-neighbor correlations. If $\hat{k}=\hat{\jmath}+1$,
Eqn. (\ref{corrRD2P}) reduces to 
\begin{eqnarray}
\langle n_{\hat{\jmath}}n_{\hat{\jmath}+1}\rangle _{t}=\left\{ 
\begin{array}{ll}
\frac{ 1-2c_{1}^{eo}(t)+c_{2}^{oo}(t)}{4} = \mathcal{C}_{1}^{oe}(t), \;
\mbox{$\hat{\jmath}$ even} &  \\ 
\frac{ 1-2c_{1}^{eo}(t)+c_{2}^{ee}(t)}{4} = \mathcal{C}_{1}^{eo}(t), \;
\mbox{$\hat{\jmath}$ odd } & 
\end{array}
\right.  \label{Cnn}
\end{eqnarray}

Again, we should emphasize that the quantities $c_{n}^{eo}(t)$, $
c_{n}^{ee}(t)$ and $c_{n}^{oo}(t)$ which appear in this section are the spin
correlations for a particular initial condition (cf. Eqn. (\ref{FINAL})), in
contrast to the more general correlations computed in Appendix B.

It is interesting to note that, generically, $\mathcal{C}_{1}^{oe}(t)\neq 
\mathcal{C}_{1}^{eo}(t)$. Of course, after a little thought this becomes
less surprising, since $\langle n_{\hat{\jmath}}n_{\hat{\jmath}+1}\rangle
_{t}$ involves the $4$-spin correlation $\langle \sigma _{j-1}\sigma
_{j}\sigma _{j}\sigma _{j+1}\rangle _{t}=\langle \sigma _{j-1}\sigma
_{j+1}\rangle _{t}$. So, if $\hat{\jmath}$ is odd (even), both $j-1$ and $
j+1 $ are even (odd), leading to a contribution of $c_{2}^{ee}(t)$ vs. $
c_{2}^{oo}(t)$, respectively.

For the general case, when $\hat{k}$ and $\hat{\jmath}$ are not nearest
neighbors, this difference between $\mathcal{C}_{\hat{k}-\hat{\jmath}
}^{eo}(t)$ and $\mathcal{C}_{\hat{k}-\hat{\jmath}}^{oe}(t)$ does not
persist. If $\hat{k}$ is even and $\hat{\jmath}$ is odd, we find: 
\begin{eqnarray}
\mathcal{C}_{\hat{k}-\hat{\jmath}}^{eo}(t) &=&\frac{1}{4}\left[ \left(
1-c_{1}^{eo}(t)\right) ^{2}-[c_{k-j}^{eo}(t)]^{2}\right]   \nonumber \\
&+&\frac{1}{4}c_{k-j+1}^{ee}(t)c_{k-j-1}^{oo}(t)  \label{Ceo}
\end{eqnarray}
and for $\hat{k}$ odd and $\hat{\jmath}$ even, one obtains 
\begin{eqnarray}
\mathcal{C}_{\hat{k}-\hat{\jmath}}^{oe}(t) &=&\frac{1}{4}\left[ \left(
1-c_{1}^{eo}(t)\right) ^{2}-[c_{k-j}^{eo}(t)]^{2}\right]   \nonumber \\
&+&\frac{1}{4}c_{k-j+1}^{oo}(t)c_{k-j-1}^{ee}(t)  \label{Coe}
\end{eqnarray}
Thanks to the simple relation between even-even and odd-odd spin
correlations, Eqn. (\ref{FINAL}), the two right-hand sides are now identical.

A similar line of reasoning shows that $\mathcal{C}_{\hat{k}-\hat{\jmath}
}^{ee}(t)=\mathcal{C}_{\hat{k}-\hat{\jmath}}^{oo}(t)$ for arbitrary
separation $\hat{k}-\hat{\jmath}$. Invoking the two-spin correlations again,
we may write

\begin{eqnarray}
&& \mathcal{C}_{\hat{k}-\hat{\jmath}}^{ee}(t)=\mathcal{C}_{\hat{k}-\hat{
\jmath} }^{oo}(t) \\
&&=\frac{1}{4}\left[ \left( 1-c_{1}^{eo}(t)\right)
^{2}-c_{k-j}^{oo}(t)c_{k-j}^{ee}(t)+c_{k-j+1}^{eo}(t)c_{k-j-1}^{eo}(t)\right]
\nonumber  \label{Cee}
\end{eqnarray}

In the following, we discuss the consequences of these results. We first
consider the steady state. Recalling our previous analysis of the spin
correlations, Eqn. (\ref{cinf}), the stationary limit of the density-density
correlations becomes very simple:\ Provided $\hat{k}-\hat{\jmath}>1$, we
find $\mathcal{C}_{\hat{k}-\hat{\jmath}}^{\mu (\hat{k}),\mu (\hat{\jmath}
)}(\infty )= \frac{1}{4} \left( 1-c_{1}^{eo}(\infty )\right) ^{2}=\rho
^{2}(\infty )$. In other words, the two-point correlations of\textit{\ }
\emph{\ non-nearest-neighbor} sites factorize into one-point functions,
independent of parity. This kind of mean-field-like behavior is typical of 
\textit{free fermion} systems \cite{FF,MM}. However, the nonequilibrium
nature of this model still imposes its signature. Turning to the \emph{
nearest-neighbor} correlations, we find that this simple factorization no
longer holds -- except in the special case where $\gamma _{e}=\gamma _{o}$.
More specifically, we find 
\begin{equation}
\mathcal{C}_{1}^{oe}(\infty )=\rho ^{2}(\infty )-\left( \gamma _{e}^2-\gamma
_{o}^2\right) \left(\frac{\omega}{4\alpha}\right)^2  \label{C_oe_ss}
\end{equation}
and 
\begin{equation}
\mathcal{C}_{1}^{eo}(\infty )=\rho ^{2}(\infty )+\left( \gamma _{e}^2-\gamma
_{o}^2\right) \left(\frac{\omega }{4\alpha}\right)^2  \label{C_eo_ss}
\end{equation}
Considering, e.g., $0<\gamma _{o}<\gamma _{e}$, we find that $\mathcal{C}
_{1}^{eo}(\infty )$ is enhanced over the mean-field result while $\mathcal{C}
_{1}^{oe}(\infty )$ is suppressed. This can be understood easily: Since $
\gamma _{o}<\gamma _{e}$ implies $T_{e}<T_{o}$, energetically costly spin
flips occur more frequently on odd sites $j$, creating a particle pair on
the nearest-neighbor dual sites ($\hat{\jmath}+1,\hat{\jmath}$). Clearly,
these sites form an ($e,o$) pair. Moreover, the rate for pair annihilation
is lower on ($e,o$) sites. Hence, particle pairs are more likely to reside
on ($e,o$) than on ($o,e$) sites. This also implies that ($e,o$) sites act
as net particle sources, while ($o,e$) sites function as sinks \cite{Beate2}
. Not surprisingly, therefore, we find $\mathcal{C}_{1}^{eo}(\infty )> 
\mathcal{C}_{1}^{oe}(\infty )$. By virtue of this reasoning, it is also
immediately apparent that this difference can only persist for
nearest-neighbor correlations. The same argument holds for $\gamma
_{o}<0<\gamma _{e}$.

A direct consequence of $\mathcal{C}_{1}^{oe}(t)\neq \mathcal{C}_{1}^{eo}(t)$
is the presence of a peculiar directional preference in the RDS. If we
consider a particle on site $\hat{\jmath}$, we can ask for the average rate, 
$\mathcal{R}_{\hat{\jmath}}(t)$, with which it will jump to the left (i.e.,
to site $\hat{\jmath}-1$) vs to the right, defined as $\mathcal{R}_{\hat{
\jmath}}(t)\equiv \frac{1}{2}\langle n_{\hat{\jmath}}(1-n_{\hat{\jmath}
+1})-n_{\hat{\jmath}}(1-n_{\hat{\jmath}-1})\rangle _{t}$. Here, the first
(second)\ term is the average rate for a particle on site $\hat{\jmath}$ to
jump to site $\hat{\jmath}+1$ ($\hat{\jmath}-1$). In our case, one might
expect this difference to vanish since neither bulk rates nor boundaries
impose a directional bias. Moreover, to avoid a potential bias at $t=0$, we
start from a translationally invariant initial condition with $\rho (0)=1/2$
. Yet, since $\mathcal{R}_{\hat{\jmath}}(t)\propto $ $\mathcal{C}
_{1}^{oe}(t)-\mathcal{C}_{1}^{eo}(t)$, it is manifestly nonzero. Explicitly,
we find: 
\begin{eqnarray}  \label{dir}
\mathcal{R}_{\hat{\jmath}}(t)=\left\{ 
\begin{array}{ll}
\frac{1}{8}\left[ 1-\frac{\gamma _{o}}{\gamma _{e}}\right] c_{2}^{ee}(t) & 
\mbox{,
$\hat{\jmath}$ even  } \\ 
\frac{1}{8}\left[ \frac{\gamma _{o}}{\gamma _{e}}-1\right] c_{2}^{ee}(t) & 
\mbox{,
$\hat{\jmath}$ odd }.
\end{array}
\right.
\end{eqnarray}
which even persists in the steady state:

\[
\mathcal{R}_{\hat{\jmath}}(\infty )=\left\{ 
\begin{array}{ll}
(\gamma_e^2 - \gamma_o^2)\,\left(\frac{\omega }{4\alpha}\right)^2 & \mbox{,
$\hat{\jmath}$ even  } \\ 
(\gamma_o^2 - \gamma_e^2)\,\left(\frac{\omega }{4\alpha}\right)^2 & \mbox{,
$\hat{\jmath}$ odd }.
\end{array}
\right. 
\]
Specifically, for $\gamma _{o}<\gamma _{e}$, particles on an even (odd)\
site jump preferentially to the right (left). Of course, this directional
preference vanishes as soon as $\gamma _{e}=\gamma _{o}$. Moreover, even
when it is nonzero, it does not generate a mass current. Counting the\emph{\
net} flow of particles between sites $\hat{\jmath}$ and $\hat{\jmath}+1$,
the natural definition of such a current is $\mathcal{J}_{\hat{\jmath}}(t)= 
\frac{1}{2}\langle (1-n_{\hat{\jmath}})n_{\hat{\jmath}+1}-n_{\hat{\jmath}
}(1-n_{\hat{\jmath}+1})\rangle _{t}$. Clearly, this expression reduces to a
density difference which vanishes for all times provided the initial
condition is homogeneous. For inhomogeneous initial condition, $\mathcal{J}_{
\hat{\jmath}}(t)$ exhibits nonzero transients for finite times but decays as 
$t\rightarrow \infty $.

Let us conclude this section with a few brief remarks about the validity of
the mean-field approximation for this system. We already noted that it does
not predict the nearest-neighbor correlations correctly, except in the
special case $\gamma_{e}=\gamma_{o}$. We now show that it also generically
misses the stationary density.

We begin by recalling Eqn. (\ref{EOMconc}). Seeking a translationally
invariant (modulo $2$) solution with $\rho _{2\hat{\jmath}}(t)=\rho _{e}(t)$,
$\rho _{2\hat{\jmath}+1}(t)=\rho _{o}(t)$ for all $\hat{\jmath}$, the
mean-field approximation corresponds to truncating two-point functions: $
\left\langle n_{\hat{\jmath}}(t)n_{\hat{\jmath}\pm 1}(t)\right\rangle \simeq
\rho _{e}(t)\rho _{o}(t)$. Starting from a uniform initial density $\rho (0)$
, we find $\rho _{e}(t)=\rho _{o}(t)\equiv \rho _{MF}(t)$, with 
\begin{eqnarray}
\rho _{MF}(t) &=&\frac{\rho _{p}\left[ \rho (0)-\rho _{m}\right] -\rho _{m} 
\left[ \rho (0)-\rho _{p}\right] e^{-t\sqrt{4-(\gamma _{e}+\gamma _{o})^{2}}
} }{\rho (0)-\rho _{m}+\left[ \rho (0)-\rho _{p}\right] e^{-t\sqrt{4-(\gamma
_{e}+\gamma _{o})^{2}}}}  \nonumber \\
&\simeq &\rho _{p}-\rho _{m}\,
\left(\frac{\rho (0)-\rho _{p}}{\rho (0)-\rho _{m}}\right)
\,e^{-t\sqrt{4-(\gamma _{e}+\gamma _{o})^{2}}},  \label{rho_MF}
\end{eqnarray}
where 
\[
\rho _{p,m}=\frac{1}{2}\left[ 1-\frac{2}{\gamma _{e}+\gamma _{o}}\pm \frac{
\sqrt{4-(\gamma _{e}+\gamma _{o})^{2}}}{\gamma _{e}+\gamma _{o}}\right] 
\]
The stationary limit is clearly $\rho _{MF}(\infty )=\rho _{p}$ which
differs from our exact result, Eqn. (\ref{densstat}), except if $\gamma
_{e}=\gamma _{o}$. In other words, the remarkable accuracy \cite{Mendoca,MM}
of the mean-field approximation for the stationary state of the uniform
system ($\gamma _{e}=\gamma _{o}$) appears to be an ``accident'' due to 
the fact that when rates are uniform the steady state is a product measure. 
We also
note that the exact relaxation time to the steady state, $\tau _{exact}=[2- 
\sqrt{\gamma _{e}\gamma _{o}}]^{-1}$, does not coincide with the mean-field
prediction, $\tau _{MF}=[\sqrt{4-(\gamma _{e}+\gamma _{o})^{2}}]^{-1}$. For
such dynamic quantities, the exact and the approximate results differ for
any choice of $\gamma _{e}$ and $\gamma _{o}$. In particular, \ the
mean-field theory always predicts an exponential decay to the steady state,
completely missing the possibility of oscillatory behavior.

\section{Solvability and relationship with free fermion systems}

The crucial ingredient for the solvability of the KISC is the quadratic spin
dependence of its Glauber-like kinetics. Thanks to this simple form, the
hierarchy of equations for the correlation functions is closed:\ to solve
the equations for the $N$-spin correlation functions, one needs to know only 
$m$-point correlations with $m\leq n$.

In RDS language, the dynamics of the particles can be rewritten as a \textit{%
\ free fermion} model, by defining a suitable quadratic (but non-Hermitian)
``stochastic Hamiltonian''. Following standard methods \cite%
{Felderhof,Schutzrev,FF,MM}, we can rewrite the master equation for the RDS
as a formal imaginary-time Schr\"{o}dinger equation: $(d/dt)|P(t)\rangle
=-H|P(t)\rangle $. The Hamiltonian $H$ is constructed by associating the
usual Pauli matrices $\sigma _{\hat{\jmath}}^{-}$ ($\sigma _{\hat{\jmath}
}^{+}$) with the creation (annihilation) of a particle at site $\hat{\jmath}$
: 
\begin{widetext}
\begin{eqnarray}
\label{Ham}
&&-2H=\sum_{\hat{\jmath} \, even}\left[
\sigma_{\hat{\jmath}}^{+} \sigma_{\hat{\jmath}+1}^{-} + \sigma_{\hat{\jmath}}^{-} 
\sigma_{\hat{\jmath}+1}^{+} + (1+\gamma_e)\sigma_{\hat{\jmath}}^{+} \sigma_{\hat{\jmath}+1}^{+} 
+ (1-\gamma_e)\sigma_{\hat{\jmath}}^{-} \sigma_{\hat{\jmath}+1}^{-}
-\gamma_e (\sigma_{\hat{\jmath}}^- \sigma_{\hat{\jmath}}^+ 
+ \sigma_{\hat{\jmath}+1}^- \sigma_{\hat{\jmath}+1}^+) -(1-\gamma_e)
\right] \nonumber\\ &+&
\sum_{\hat{\jmath} \, odd}\left[
\sigma_{\hat{\jmath}}^{+} \sigma_{\hat{\jmath}+1}^{-} + \sigma_{\hat{\jmath}}^{-} 
\sigma_{\hat{\jmath}+1}^{+} + (1+\gamma_o)\sigma_{\hat{\jmath}}^{+} \sigma_{\hat{\jmath}+1}^{+}
+ (1-\gamma_o)\sigma_{\hat{\jmath}}^{-} \sigma_{\hat{\jmath}+1}^{-}
-\gamma_o(\sigma_{\hat{\jmath}}^- \sigma_{\hat{\jmath}}^+ 
+ \sigma_{\hat{\jmath}+1}^- \sigma_{\hat{\jmath}+1}^+) -(1-\gamma_o)
\right]
\end{eqnarray}
\end{widetext}

The key to the solvability of this Schr\"{o}dinger equation lies in the
bilinear dependence of the Hamiltonian on the Pauli matrices. This is due to
the fact that the spin-flip rates (\ref{rates}) implicitly fulfill the 
\textit{free fermion} constraint \cite{Schutzrev,FF,MM}. In RDS language,
this condition requires that the sum of the particle diffusion rates equal
the sum of the (local)\ annihilation and creation rates, i.e., $
1/2+1/2=(1+\gamma _{j})/2+(1-\gamma _{j})/2$ with $j\in \{e,o\}$ in our
case. If this relation is violated, $H$ includes quartic terms, of the form $
\sum_{\hat{\jmath}}\sigma _{\hat{\jmath}}^{-}\sigma _{\hat{\jmath}
}^{+}\sigma _{\hat{\jmath}+1}^{-}\sigma _{\hat{\jmath}+1}^{+}$, and the
associated RDS\ can no longer be solved exactly. It can, of course, be
simulated, and for those cases investigated so far, it appears that the
quartic terms are irrelevant for the long-time dynamics \cite{FF,Mendoca}. It
is also worth noting that the free fermion constraint is not particularly
artificial: the simplest models for photogenerated solitons in MX chain
compounds satisfy it quite naturally \cite{exp1}. 

Here we decided to invoke  
generating function techniques instead of diagonalizing (\ref{Ham}).
In our view this is the most convenient and systematic approach to solve {\it 
simultaneously} both the KISC and RDS, for two reasons. 
First, the free-fermion approach requires various technical steps
({\it e.g.} introduction of so-called pseudo-fermion operators and a
Bogoliubov-like transformation) which make the general 
treatment rather involved, especially for arbitrary initial conditions \cite{FF,MM} . 
Further, the diagonalization of (\ref{Ham}) yields only correlation 
functions with an {\it even} number of spins (see Section III and IV); 
the calculation of correlations involving an {\it odd} number of spins requires a 
dual transformation of (\ref{Ham}) into a new stochastic Hamiltonian which must 
also be diagonalized \cite{Santos}.

Let us also mention that damped oscillatory decay has been observed before
in certain reaction-diffusion models \cite{Schutzrev}. However, those
models, and hence the physical mechanisms leading to the oscillations, are
completely different from ours. As an example, a diffusion-limited fusion
model \cite{Schutzrev} is defined by three processes: (\emph{i}) biased
diffusion: $A\emptyset \rightarrow \emptyset A$ with rate $D(1+\eta )$, $
\emptyset A\rightarrow A\emptyset $ with rate $D(1-\eta )$ (with $0<\eta
\leq D$); (\emph{ii}) biased fusion: $AA\rightarrow \emptyset A$ with rate $
D(1+2\eta )$, $AA\rightarrow A\emptyset $ with rate $D(1-2\eta )$; 
and (\emph{iii}) homogeneous pair production: $\emptyset \emptyset \rightarrow AA$
, with rate $D$. With this special choice of rates, the equation of motion
for the density closes and becomes solvable. In order to observe oscillatory
decay of the particle density, the initial condition must be \emph{
inhomogeneous}. For a homogeneous initial condition, the density decays
exponentially. In contrast, the equation in our reaction-diffusion model
does not close, and the oscillatory behavior is generic: it occurs for any
initial condition, inhomogeneous or not.

\section{Conclusions.}

To summarize, we have presented a full exact solution for the dynamics of a
non-equilibrium Ising spin chain, with arbitrary initial condition. The
model is characterized by a generalization of Glauber dynamics: spins on
even/odd sites are coupled to alternating temperatures, $T_{e}$ and $T_{o}$.
We obtain all correlation functions from a generating functional. As an
illustration, we have discussed the equal-time and the two-time spin-spin
correlation functions.

Identifying domain walls in the spin system with particles on the dual
lattice, the model can also be interpreted as a reaction-diffusion system.
Particles are created and annihilated in pairs; the rates for these
processes alternate from even to odd sites. This mapping opens up an
interesting extension of parameter space: while negative temperatures are
unphysical for the spin chain, the corresponding rates are perfectly natural
in the context of the RDS. By expressing particle-particle correlations as
superpositions of spin-spin correlation functions, the RDS becomes exactly
soluble. This is not entirely trivial since the BBGKY \cite{BBGKY} hierarchy
for the RDS is \emph{not} closed: its solution is far from obvious unless
one recognizes the connection to the spin chain.

For $0<\gamma _{e}\gamma _{o}$, energetically favorable spin flips always
dominate over unfavorable ones, irrespective of whether they occur on even
or odd sites. In RDS language, pair annihilation is always more probable
than pair creation. As a consequence, we find that all quantities decay
exponentially to their steady-state values. In contrast, for $\gamma
_{e}\gamma _{o}<0$, we observe (damped) \emph{oscillatory} behavior. Its
origin can be traced to a\emph{\ competition} of pair creation and
annihilation on even vs odd sites on the original lattice: If, say, $\gamma
_{o}<0$, then pair creation dominates over annihilation on odd sites while
the relation is reversed on the even sites. Hence, a given initial particle
density may first decrease, due to annihilation processes and then recover,
as the available empty sites are (partially)\ filled again by the strong
creation process, and so on, until the stationary density is reached.

Remarkably, even in the absence of any bias in the rates, boundary or
initial conditions, particles still ``know'' the difference between right
and left: For, e.g., $\gamma _{o}<\gamma _{e}$, particles on even (odd)
sites jump preferentially to the right (left). Even though this directional
preference does not lead to a systematic particle current, it is still
somewhat surprising. However, once we recall that particles are most often
created (annihilated) on pairs of neighboring sites, with the odd site on
the left (right), we recognize that the directional preference is simply a
response to this density gradient.

Since exact solutions, especially of a full nonequilibrium dynamics are
rare, we hope that our model can serve as a testing ground for various
generalizations or approximations. The features reported here -- exponential
decays, damped oscillations, and directional preference -- should be generic
for a whole class of genuine out-of-equilibrium models. Moreover, they
should be experimentally observable in MX chain compounds exposed to
spatially modulated laser light.

\vspace{0.3cm}

\textbf{Acknowledgments:} It is a pleasure to acknowledge fruitful
discussions with I. Georgiev, H. Hilhorst, J.R. Heflin, and U.C. T\"{a}uber.
MM acknowledges the support of Swiss NSF Fellowship No. 81EL-68473. This
work was also partially supported by US NSF DMR-0088451, 0308548 and 0414122.

\begin{widetext}
\appendix

\section{The derivation of the generating function}

In this appendix we provide some details for the derivation of the generating
function (\ref{gen}), which is one of the key results of this work. We
follow Aliev's work and notation \cite{Aliev}. Aliev established that the generating 
functions of a very general class of
disordered Glauber-Ising spin chains, including our case, 
can formally be expressed in terms of two functions 
$M^{\pm}_{j,k}(t)$ and two additional quantities $W_{j,k}^{\pm}(t)$, which depend in a very
involved fashion on $M^{\pm}_{k,j}(t)$. Below, we will see that these quantities are 
closely related to physical observables, namely, the magnetization and the two-point 
correlations. 
Here, we follow Aliev by noting that the Laplace
transform of ${\hat M}^{\pm}_{j,k}$ is the inverse of an $L\times L$ band matrix $
(s+1)\openone - \frac{1}{2}U^{\pm}$. For our case, the entries
of this matrix can be taken from the rates and read explicitly ($L$
is even): 
\begin{eqnarray}  \label{matrix}
\left[(s+1)\openone -\frac{1}{2}U^{\pm}\right]_{2j-1,k}&=&(s+1)\delta_{2j-1,k}-
\frac{\gamma_o}{2}(\delta_{2j-1,k-1}+\delta_{2j-1,k+1}), \quad (1 < j\leq
L/2)  \nonumber \\
\left[(s+1)\openone -\frac{1}{2}U^{\pm}\right]_{2j,k}&=&(s+1)\delta_{2j,k}-
\frac{\gamma_e}{2}(\delta_{2j,k-1}+\delta_{2j,k+1})\quad (1\leq j<L/2)
\nonumber \\
\left[(s+1)\openone -\frac{1}{2}U^{\pm}\right]_{1,k}&=&(s+1)\delta_{1,k}-
\frac{\gamma_o}{2}(\delta_{2,k} \mp\delta_{k,L})  \nonumber \\
\left[(s+1)\openone -\frac{1}{2}U^{\pm}\right]_{L,k}&=&(s+1)\delta_{L,k}-
\frac{\gamma_o}{2}( \delta_{L-1,k} \mp \delta_{1, k})
\end{eqnarray}
Given (\ref{matrix}), it is easy to evaluate the inverse of 
$[(s+1) \openone - \frac{1}{2}U^{\pm}]_{j,k}$:  
\begin{eqnarray}
{\hat M^{\pm}}_{k,j}=\frac{1}{L}\sqrt{\frac{\gamma_k}{\gamma_j}}%
\sum_{n=1}^{L}\frac{e^{i(k-j)\phi_n^{\pm}}}{s+1-\alpha \cos{\phi_n^{\pm}}},
\end{eqnarray}
where $\phi_n^{+}=\frac{\pi(2n-1)}{L}$ and $\phi_n^{-}=\frac{2 \pi n}{L}$,
with $n=1,2,\dots, L$. In the thermodynamic limit $L\to \infty$, the two 
quantities ${\hat M^{+}}_{j,k}$ and ${\hat M^{-}}_{j,k}$ coincide whence we 
simply have 
\begin{eqnarray}  \label{contlim}
{\hat M^{\pm}}_{k,j}\to {\hat M}_{k,j}= \sqrt{\frac{\gamma_k}{\gamma_j}}
\int_{0}^{2\pi} \frac{d\phi}{2\pi}\, \frac{e^{i(k-j)\phi}}{s+1-\alpha \cos{\phi}}.
\end{eqnarray}
Taking the inverse Laplace transform of (\ref{contlim}), we recover Eqn. 
(\ref{M(t)}) for the propagator.

Since the $W_{j,k}^{\pm}(t)$ can be expressed in terms of the $M^{\pm}_{j,k}(t)$, 
we may immediately conclude that $W_{j,k}^{\pm}(t)\to W_{j,k}(t)$ as $L\to \infty$.
According to Aliev, $W_{j,k}(t)$ is simply the two-point correlation
function, $\langle \sigma_j \sigma_{k\neq j}\rangle_t$, for a particular
initial condition, namely, $m_j(0)=0$ and 
$\langle \sigma_j \sigma_{k\neq j}\rangle_0=0$. 
For our case, these correlations were given in Eqn.(\ref{two-spin}).

For readers familiar with Aliev's work \cite{Aliev}, these remarks fill in the gaps between
Aliev's formal and general analysis and the special case we are interested in here.
It follows that the generating function
of our KISC admits the compact and explicit representation of Eqn. (\ref{gen}) 
which encodes the \textit{complete} dynamics of the system.

\section{Translationally-invariant initial conditions: The spin-spin correlation functions and the particle density}

In Section III.B, we have derived an exact expression, Eqn. (\ref{2P}), for the
spin-spin correlation functions of our KISC, valid for arbitrary
initial conditions. Here, we impose a natural restriction, namely, translational
invariance, on the initial conditions. Thanks to the symmetry, 
Eqn. (\ref{2P}) simplifies considerably, as we will show now.

As we already pointed out in \cite{MZS}, for translationally invariant 
initial conditions, we only need to consider the correlations 
between spins at two even sites, two odd sites, and one even, one odd site. 
We denote these by $c_{2n}^{ee}(t)\equiv
\langle \sigma _{2\ell }\sigma _{2(\ell +n)}\rangle _{t}$, $
c_{2n}^{oo}(t)\equiv \langle \sigma _{2\ell -1}\sigma _{2\ell -1+2n}\rangle
_{t}$, $c_{2n-1}^{eo}(t)=c_{2n-1}^{oe}(t)\equiv \langle \sigma _{2\ell }\sigma _{2\ell
+2n-1}\rangle _{t}=\langle \sigma _{2\ell
+1}\sigma _{2\ell +2n}\rangle _{t}$.
Of course, there is no need to study $n<0$
cases. 
For the special case of zero initial magnetization and correlations, these correlations 
are already known \cite{MZS} and are given by Eqn. (\ref{FINAL}). 
Here, we seek their form in a more general case, starting from a homogeneous
initial condition.

Let us recall from \cite{MZS}
that in the translationally-invariant case the quantities $a_{2n}\left(
t\right) \equiv \frac{1}{2}\left[ \gamma _ec_{2n}^{oo}\left( t\right)
+\gamma _oc_{2n}^{ee}\left( t\right) \right], 
a_{2n-1}\left( t\right) \equiv  \alpha  c_{2n-1}^{eo}\left(
t\right) $, obey the following simple
equation: $ \frac d{dt}a_j=-2a_j+\alpha [a_{j-1}+a_{j+1}]\,\,,\quad j>0 $
with the initial condition $a_0\left( t\right) = \bar{\gamma}$ and 
$ \bar{\gamma}\equiv \left( \gamma_e+\gamma _o\right)/2$.
The equations of motion of the KISC \cite{MZS} also give the following 
relationships among the correlators: 
$c_{2n}^{oo}(t)=\frac{\gamma_o}{\gamma_e} c_{2n}^{ee} (t)
+\left(c_{2n}^{oo}(0)-\frac{\gamma_o}{\gamma_e} c_{2n}^{ee} (0)
\right)e^{-2t}.$

The explicit expressions for the
correlators follow from (\ref{2P}), or by the methods of images directly from  $a_{n}(t)=a_{n}(\infty) +e^{-2t}\sum_{m\geq
0}[a_{m}(0)-a_{m}(\infty)] \, \left\{ I_{n-m}(2\alpha t) - I_{n+m}(2\alpha
t) \right\}$ where $a_{n}(\infty)={\bar \gamma}\, \omega^k$ and $n\geq 0$. From the
definitions of $a_n$, we immediately infer:

\begin{eqnarray}  \label{cne}
c_{2n}^{ee} (t)&=& \frac{\gamma_e}{\gamma_o} c_{2n}^{oo}(t)-\left( \frac{
\gamma_e}{\gamma_o}c_{2n}^{oo}(0) -c_{2n}^{ee}(0) \right) e^{-2t}  \nonumber
\\
&=& \frac{a_{2n}(\infty)}{\gamma_o}+ \frac{e^{-2t}}{\gamma_o}\sum_{m\geq
0}[a_{2m}(0)-a_{2m}(\infty)] \, \left\{ I_{2(n-m)}(2\alpha t) -
I_{2(n+m)}(2\alpha t) \right\}  \nonumber \\
&+&\frac{\alpha e^{-2t}}{\gamma_o} \,%
\sum_{m>0}[c_{2m-1}^{eo}(0)-c_{2m-1}^{eo}(\infty)] \, \left\{
I_{2(n-m)+1}(2\alpha t) - I_{2(n+m)-1}(2\alpha t)\right\}  \nonumber \\
&-&\frac{e^{-2t}}{2\gamma_o}\, \left( \gamma_e c_{2n}^{oo}(0) - \gamma_o
c_{2n}^{ee}(0) \right).
\end{eqnarray}

Following the same steps for $c_{2n-1}^{eo}=c_{2n-1}^{oe}$, 
we obtain, for $(n>0)$:

\begin{eqnarray}  \label{cneo}
c_{2n-1}^{eo}(t)&=&c_{2n-1}^{eo}(\infty)+
e^{-2t}\sum_{m>0}[c_{2m-1}^{eo}(0)-c_{2m-1}^{eo}(\infty)] \, \left\{
I_{2(n-m)}(2\alpha t) - I_{2(n+m-1)}(2\alpha t)\right\}  \nonumber \\
&+&\frac{e^{-2t}}{\alpha}\sum_{m>0}[a_{2m}(0)- a_{2m}(\infty)] \, \left\{
I_{2(n-m)-1}(2\alpha t) - I_{2(n+m)- 1}(2\alpha t) \right\}.
\end{eqnarray}

The expressions (\ref{cne})-(\ref{cneo}) illustrate that the time-dependence
of the spin-spin correlation function depends non-trivially on the initial
condition, and we may therefore anticipate \textit{non-universal} behavior. 
Of course, when $\gamma_e=\gamma_o$, the expressions 
(\ref{cne}), (\ref{cneo}) coincide with those obtained by Glauber \cite
{Glauber}. 

An interesting situation occurs when, say,
$\gamma_o$ is negative while $0<\gamma_e \leq 1$, so that 
$\alpha=i|\alpha|$.
Then, we have
 $I_{2n}(2i|\alpha|t)=(-1)^{n}J_{2n}(2|\alpha|t)$ and $I_{2n\pm
1}(2i|\alpha|t)=\pm i\, (-1)^{n}J_{2n\pm 1}(2|\alpha|t)$, where $
J_{n}(x)\equiv \int_{0}^{\pi}\frac{dq}{\pi}\cos\left( x\sin{q} -nq\right)$
is the Bessel function of first kind \cite{Abramowitz}. Further, when $
\alpha=i|\alpha|$, the expressions (\ref{cne}),(\ref{cneo})  become:

\begin{eqnarray}  \label{cneTneg}
c_{2n}^{ee} (t)&=&\frac{\gamma_e}{\gamma_o} c_{2n}^{oo}(t)-\left( \frac{
\gamma_e}{\gamma_o}c_{2n}^{oo}(0) -c_{2n}^{ee}(0) \right) e^{-2t}  \nonumber
\\
&=&-\frac{a_{2n}(\infty)}{|\gamma_o|}- \frac{e^{-2t}}{|\gamma_o|}\sum_{m\geq
0}[a_{2m}(0)-a_{2m}(\infty)] \,(-1)^{n+m}\, \left\{ J_{2(n-m)}(2|\alpha| t)
- J_{2(n+m)}(2|\alpha|t) \right\}  \nonumber \\
&+&\frac{|\alpha| e^{-2t}}{|\gamma_o|} \,\sum_{m>0}
[c_{2m-1}^{eo}(0)-c_{2m-1}^{eo}(\infty)] \,(-1)^{n+m}\, \, \left\{
J_{2(n-m)+1}(2|\alpha| t) + J_{2(n+m)-1}(2|\alpha| t)\right\}  \nonumber \\
&-&\frac{e^{-2t}}{2\gamma_o}\, \left( \gamma_e c_{2n}^{oo}(0) - \gamma_o
c_{2n}^{ee}(0) \right).
\end{eqnarray}
\begin{eqnarray}  \label{cneoTneg}
c_{2n-1}^{eo}(t)&=&c_{2n-1}^{eo}(\infty)+e^{-2t}\sum_{m>0}[c_{2m-1}^{eo}(0)-
c_{2m-1}^{eo}(\infty)] \, (-1)^{(n+m)}\left\{ J_{2(n-m)}(2|\alpha| t) +
J_{2(n+m-1)}(2|\alpha| t)\right\}  \nonumber \\
&-&\frac{e^{-2t}}{|\alpha|}\sum_{m>0}[a_{2m}(0)- a_{2m}(\infty)] \,
(-1)^{(n+m)} \left\{ J_{2(n+m)-1}(2|\alpha| t) - J_{2(n-m)-1}(2|\alpha| t)
\right\}.
\end{eqnarray}
In the long-time limit, these expressions exhibit a damped oscillatory approach to the
stationary state.

We now turn to the equivalent RDS. As we have seen in Section IV.A , the density of particles
is related to the nearest-neighbor spin correlations, $c_{1}^{eo}$, according to 
$\rho(t)=\frac{1}{2}(1-c_1^{eo}(t))$.
Here, our goal is to determine the long-time behavior of this density for a homogeneous (but otherwise
arbitrary) initial concentration of particles $\rho(0)$.
In this respect, the expressions (\ref{cneo}) and (\ref{cneoTneg}) are not very practical as they involve infinite
sums of Bessel functions. At this point, for further convenience, it is useful to introduce four auxiliary functions 
defined as follows (with $0\leq \lambda\leq 1$):

\begin{eqnarray}
\label{F1}
F_1(\lambda,t) &\equiv &\sum_{m>0}\lambda ^{2m-1}\,e^{-2t}\left\{
I_{2(m-1)}(2\alpha t)-I_{2m}(2\alpha t)\right\}  
=2\lambda(1+\lambda^2) \int_{0}^{\pi }\frac{dq}{\pi }\,\frac{e^{-2t(1-\alpha \cos {q})}
\sin^2{q}}{1+\lambda
^{4}-2\lambda ^{2}\cos {2q}} \\
\label{F2}
F_{2}(\lambda,t) &\equiv &\sum_{m>0}\lambda ^{2m}\,e^{-2t}\left\{
I_{2m-1}(2\alpha t)-I_{2m+1}(2\alpha t)\right\} =2\lambda
^{2}\int_{0}^{\pi }\frac{dq}{\pi }\,e^{-2t(1-\alpha \cos {q})}\,\left[ \frac{
\sin {2q}\sin {q}}{1+\lambda ^{4}-2\lambda ^{2}\cos {2q}}\right]\\
\label{G1}
G_{1}(\lambda,t)&\equiv& -\sum_{m>0} \lambda^{2m-1} \,
(-1)^{m} e^{-2t}\left\{J_{2(1-m)}(2|\alpha| t)+J_{2m} (2|\alpha| t)
\right\}  \nonumber \\
&=& 2\lambda(1+\lambda^2) \int_{0}^{\pi} \frac{dq}{\pi}\, e^{-2t}\,\cos{
(2|\alpha|t\sin{q})}\, \left[\frac{\cos^2{q}}{1+\lambda^4+2
\lambda^2 \cos{2q}}\right] \\
\label{G2}
G_{2}(\lambda,t)&\equiv& -\sum_{m>0}\lambda^{2m} \, (-1)^{m} e^{-2t}
\left\{J_{2m+1}(2|\alpha| t)-J_{1-2m} (2|\alpha| t) \right\} \nonumber\\
&=& 2\lambda^2  \int_{0}^{\pi} \frac{dq}{\pi}\, e^{-2t}\sin{(2|\alpha|
t\sin{q})} \left[\frac{\sin{2q}\cos{q}}{1+\lambda^4+2\lambda^2 \cos{2q}
}\right]
\end{eqnarray}

To establish these expressions, we have invoked the integral representation of the
Bessel functions \cite{Abramowitz} and the properties of geometric series.
With these functions and the help of Eqn. (\ref{dens}), the density of particles in the RDS 
model can now be recast in compact form. Two cases emerge naturally: 

\begin{itemize}
\item When $\gamma_e\gamma_o>0$ :
\begin{eqnarray}
\label{real}
\rho(t)-\rho(\infty)=\frac{{\bar \gamma}}{2\alpha}\left[
F_1(\omega ,t) +F_2(\omega ,t)
\right] -\frac{1}{2}\left[
F_1(1-2\rho(0),t) + \frac{{\bar \gamma}}{\alpha}F_2(1-2\rho(0),t)
\right] 
\end{eqnarray}
\item When $\gamma_e\gamma_o<0$ :
\begin{eqnarray}
\label{im}
\rho(t)-\rho(\infty)&=&
\frac{{\bar \gamma}}{2|\alpha|}\left[
G_2(1-2\rho(0),t) - iG_1(\omega ,t)
\right] -\frac{1}{2}\left[
\frac{{\bar \gamma}}{|\alpha|}G_2(\omega ,t) + G_1(1-2\rho(0),t) 
\right]
\end{eqnarray}
\item The case $\gamma_e\gamma_o=0$ is special and gives rise to a purely exponential time-dependence:
\begin{eqnarray}
\label{zero}
\rho(t)=\frac{{\bar \gamma}}{2}+ \left(\frac{{\bar \gamma}-2}{4} + \rho(0) \right)\,e^{-2t}
\end{eqnarray}
\end{itemize}

We now proceed with the analysis of the long-time behavior of these expressions.
Again, we first consider the case where $\gamma_e\gamma_o>0$ and then $\gamma_e\gamma_o<0$.
\begin{itemize}
\item When $\gamma_e\gamma_o>0$, the main contribution to the long-time behavior arises from the small $q$
contribution in the expression of the functions $F_1$ and $F_2$. Therefore, one may expand the integrand of $F_1$ and $F_2$ in 
Eqn. (\ref{real}). It is also essential to pay due attention to the initial condition:  
\begin{enumerate}
\item When $0<\gamma_e\gamma_o<1$ and $0<\rho(0)<1$ (also $\rho(0)\neq \rho(\infty)$), we obtain
\begin{eqnarray}
\label{genrho}
\rho(t)-\rho(\infty)\simeq  \frac{1}{4}\left[ \frac{{\bar \gamma} \omega }{\alpha(1-\omega ^2)} 
-\frac{\left(1+\frac{{\bar \gamma}}{\alpha}\right) \left\{1-2\rho(0)\right\} + 2\rho(0)^2}{\rho(0)(1-\rho(0))}
\right]\,\frac{e^{-2(1-\alpha)t}}{\alpha t \,\sqrt{\pi\alpha t}}.
\end{eqnarray}
\item When $0<\gamma_e\gamma_o<1$ and $\rho(0)=0$, we find
\begin{eqnarray}
\label{rhozero}
\rho(t)-\rho(\infty)\simeq -\left[1+\frac{{\bar \gamma}}{\alpha} -\left\{ \frac{{\bar \gamma}\omega }{\alpha(1-\omega ^2)}\right\}\frac{1}{\alpha t}\right]\, \frac{e^{-2(1-\alpha)t}}{4\sqrt{\pi\alpha t}}.
\end{eqnarray}
\item When $0<\gamma_e\gamma_o<1$ with $\rho(0)=1$, we have
\begin{eqnarray}
\label{rho1}
\rho(t)-\rho(\infty)\simeq  
\left[1-\frac{{\bar \gamma}}{\alpha} + 
\left\{\frac{{\bar \gamma}\omega }{\alpha(1-\omega ^2)}\right\}
\frac{1}{\alpha t}  \right]\, \frac{e^{-2(1-\alpha)t}}{4\sqrt{\pi\alpha t}}.
\end{eqnarray}
These results show that for $\gamma_e\gamma_o>0$, the density generically approaches its stationary value
as $\propto t^{-3/2}\,e^{-2(1-\alpha)t}$, with some nontrivial amplitude. Only if the lattice is
initially completely empty/occupied by particles, the long-time behavior is modified to $\propto t^{-1/2}\,e^{-2(1-\alpha)t}$
(provided $\frac{{\bar \gamma}}{\alpha}\neq \pm 1$).
 \item The case where $\gamma_e=\gamma_o=\gamma=\pm 1$ is {\it critical} and we can check from (\ref{real}) that one recovers the previously known results \cite{FF,MM}:
 \begin{eqnarray}
\label{rhocrit}
\rho(t)-\rho(\infty)\simeq  
\frac{\gamma}{2\sqrt{\pi t}}.
\end{eqnarray}
In this case, it is well known \cite{FF,MM} that the density of particles approaches
the steady state algebraically slowly ($\propto t^{-1/2}$). 
For $\gamma= 1$ (only pair annihilation), the stationary value is $\rho(\infty)=0$; in contrast, we find $\rho(\infty)=1$ for $\gamma= -1$
(only pair creation). We emphasize that for such a {\it critical} dynamics neither the dynamical exponent nor the amplitude of (\ref{rhocrit}) 
depend on the initial condition.  
\end{enumerate}
\item When $\gamma_e\gamma_o<0$, it is difficult to directly analyze the long-time behavior of the 
oscillating function $G_1$ and $G_2$; instead, we seek upper and lower bounds for $\lambda\neq \pm 1$. 
We observe  that the denominator of the integrand in the expressions for 
$G_1$ and $G_2$ can be bounded as follows: 
$(1-\lambda^2)^2 \leq (1+\lambda^4+2\lambda^2 \cos{2q})
\leq (1+\lambda^2)^2$. Therefore, we obtain for the auxiliary functions $G_1$ and $G_2$:
\begin{eqnarray}
\label{bound1}
\frac{\lambda}{1+\lambda^2}\,\frac{e^{-2t}\,J_1(2|\alpha|t)}{2|\alpha| t} \leq G_1(\lambda,t)\leq
\frac{\lambda (1+\lambda^2)}{(1-\lambda^2)^2}\, \frac{e^{-2t}\,J_1(2|\alpha|t)}{2|\alpha| t} \\
\label{bound2}
2\left(\frac{\lambda}{1+\lambda^2}\right)^2\, \frac{e^{-2t}\,J_2(2|\alpha|t)}{2|\alpha| t} \leq G_2(\lambda,t)\leq
2\left(\frac{\lambda}{1-\lambda^2}\right)^2\, \frac{e^{-2t}\,J_2(2|\alpha|t)}{2|\alpha| t}.
\end{eqnarray}
If $\lambda=\pm 1$, one has the exact expressions:
$
G_1(1,t)=-G_1(-1,t)=e^{-2t}\, J_{0}(2|\alpha|t); \;
G_2(1,t)=G_2(-1,t)=\frac{e^{-2t}}{2|\alpha|t}\, J_{1}(2|\alpha|t).$
At long times and for finite $n$, $e^{-2t}J_{n}(2|\alpha|t)\simeq \frac{e^{-2t}}{\sqrt{\pi|\alpha|t}}\, \cos{\left(
2|\alpha|t -\frac{\pi}{4}(2n+1)
\right)}$ and therefore the upper and lower bounds in Eqns. (\ref{bound1}) and (\ref{bound2}) 
display the same time-dependence. With the help of Eqn.(\ref{im}), we thus deduce:
\begin{enumerate}
\item  When $0<\rho(0)<1$ (and obviously $\rho(0)\neq \rho(\infty)$),
\begin{eqnarray}
\rho(t)-\rho(\infty)\simeq t^{-3/2}e^{-2t}\,\left(
A\cos{\left(2|\alpha|t + \frac{\pi}{4}\right)} + B\cos{\left(2|\alpha|t - \frac{\pi}{4}\right)}
\right),
\end{eqnarray}
where $A$ and $B$ are some amplitudes depending nontrivially on all the parameters of the system and on the initial density.
\item When $\rho(0)=1$, we obtain an explicit expression for the long-time behavior of the density:
\begin{eqnarray}
\rho(t)-\rho(\infty)\simeq \frac{e^{-2t}}{2}J_0(2|\alpha|t)\simeq
\frac{e^{-2t}}{2\sqrt{\pi |\alpha|t }}\,\cos{\left(2|\alpha|t - \frac{\pi}{4}\right)}.
\end{eqnarray}
\item When $\rho(0)=0$, we also have an explicit expression for the long-time behavior of the density:
\begin{eqnarray}
\rho(t)-\rho(\infty)\simeq -\frac{e^{-2t}}{2}J_0(2|\alpha|t)\simeq
\frac{e^{-2t}}{2\sqrt{\pi |\alpha|t }}\,\cos{\left(2|\alpha|t + \frac{3\pi}{4}\right)}.
\end{eqnarray}
\end{enumerate}
These results show that, for $\gamma_e\gamma_o<0$, the density displays oscillations which are damped by a factor
$t^{-\beta}e^{-2t}$, where $\beta=3/2$ for generic initial densities $\rho(0)$, with two exceptions: we have 
$\beta=1/2$ if the system is initially completely empty or occupied.

Of course, following the same approach one would be able to compute every $n-$point correlation functions, for both, the KISC and RDS. 
While perfectly straightforward, these computations become rather tedious.
\end{itemize}
\end{widetext}

\end{document}